\documentclass{article}

\usepackage{epsf,hyperref}
\usepackage{amssymb,ComplexSystems}
\usepackage{graphicx}
\newcommand{\0}{\textbf{0}}
\newcommand{\1}{\textbf{1}}
\newcommand{\?}{\textbf{?}}

\begin{document}

\title{Cooperation as well as learning:\\
A commentary on `How learning can guide evolution' by Hinton and Nowlan.%
%
}

\author{\authname{Conor Houghton}\\[2pt] 
\authadd{Intelligent Systems Laboratory, University of Bristol}\\
\authadd{Michael Ventris Building}\\
\authadd{Bristol, BS8 1UB, United Kingdom}\\
\and
}

%
%
\markboth{Conor Houghton}
{Cooperation as well as learning} 

\maketitle
\begin{abstract}
According to the Baldwin Effect learning can guide evolution. This does not suppose that information about what has been learned is transferred back into the genetic code: in the Baldwin Effect complex multi-gene characteristics are discovered through learning but acquired through standard selectionist evolutionary processes. Learning serves to improve the search by giving value to a partial, and otherwise useless, subset of the required genes.  An elegant and concrete treatment of the Baldwin Effect is given in `How learning can guide evolution', a 1987 paper by G.~E. Hinton and S.~J. Nowlan. This includes a simple, but revealing, simulation illustrating the effect. As a commentary on that paper, a similar simulation is used here to demonstrated that cooperation can also guide evolution. Like learning, cooperation has a clear benefit to survival, but what is proposed here is small addition: that cooperation, like learning in the Baldwin Effect, can also allow complex characteristic to be discovered and acquired much faster than they otherwise would. This suggests an additional benefit of social behavior and suggests that social animals have a broader evolutionary path towards some complex adaptations. 
\end{abstract}



\section{Introduction}

Many animals cooperate and, when they do, this produces an ecological benefit. It is proposed here that cooperation can also guide evolution by supporting the discovery of complex characteristics. This is akin to the Baldwin Effect. The Baldwin Effect \cite{Baldwin1896,Morgan1896,Waddington1942} describes the way learning can guide evolution by easing its path towards adaptations that would otherwise require too many non-beneficial changes to make discovery likely. The effect is of particular interest because of the role it might play in the evolution of mind and of language; \cite{PinkerBloom1990} for example argues that the Baldwin Effect may be important in the evolution of language. A good introduction to this debate is found in \cite{WeberDepew2003}. The focus of this commentary is on the description of the Baldwin Effect given in \cite{HintonNowlan1987} and reviewed in \cite{Smith1987}. Here, in this commentary, I propose that a similar argument to one in \cite{HintonNowlan1987} suggests that cooperation can also guide evolution. To this end this commentary follows \cite{HintonNowlan1987} in many respects, in particular, it uses an simulation very similar to the illustrative simulation provided in \cite{HintonNowlan1987}, but, in this case, adjusted so that it models cooperation rather than learning.

The Baldwin Effect and the proposal here are best introduced by summarizing the simulation given in \cite{HintonNowlan1987}. A collection of 1000 agents is modeled. Each agent has a genotype composed of 20 genes, which may, in this example, have values \1 or \0. Each gene codes for one property of the agent which may in turn have one of two valences, again marked \1 or \0. The genotype only gives a benefit if all the properties are \1 and an agent of this sort, with all 20 properties set to \1 will be called fit. With $2^{20}$ possible genotypes the evolutionary discovery of this beneficial configuration is daunting: there is no advantage to acquiring genes with value \1 unless all genes have that value and so evolution can only arrive at this configuration by stumbling upon it.

\begin{table}
    \centering
    \begin{tabular}{l|cccccc}
         genes&\1&\0&\?&\ldots&\1&?  \\
         properties $t=1$&\1&\0&\0&\ldots&\1&\1\\
         properties $t=2$&\1&\0&\1&\ldots&\1&\1\\
         properties $t=3$&\1&\0&\0&\ldots&\1&\0
    \end{tabular}
    \caption{\textbf{Genes and properties} This illustrates the relationship between genes and properties. Genes with values \1 and \0 determined the corresponding properties, whereas the properties corresponding to value \? are picked randomly at each time step. The \0's in the genotype mean that the properties will never all have value \1 in this case, so fitness will never be attained but a genotype with no zeros is much more likely to occur when there are \?{}s than in the case without learning.}
    \label{tab:learning}
\end{table}

Learning is now added by allowing the genes an additional value marked \?. The properties corresponding to the \? genes can be either \1 or \0. In this simple simulation there is no actual learning strategy, instead the valence for the property corresponding to a \? gene is decided randomly at each time step, as illustrated in Table~\ref{tab:learning}. Each generation survives for 1000 time steps, at any time step any fit agent stops learning and if frozen in the fit state, while all the unfit agents continue to `learn' by randomly resetting any property whose gene has the value \?. At the end of each generation a new set of 1000 agents is produced by breeding; the probability of an agent being selected as a parent depends on how early it became fit. With this in place the \0 gene disappears quickly from the population: initially the genes are set so they have a 0.5 probability of being \? and 0.25 for each of \1 and \0; by about 20 generations the \0 gene has disappeared and the genes are split between \? and \1 with a ratio of about 0.45 to 0.55; the distribution does not appear to change after this.

\section{Cooperation}

Instead of learning, cooperation is now considered. In the spirit of \cite{HintonNowlan1987} an extremely simple model is used. In the simulation there are 1024 agents, as before each has a genotype with 20 genes, each taking only the values \1 and \0. At each time step, agents are grouped into groups of four. A group is considered fit if for each of the 20 properties at least one of the four agents has the \1 gene. A fit group is frozen, but the agents in unfit groups are regrouped randomly at the start of the next time step. This repeats for 20 time steps.

After 20 time steps a new set of agents is generated by a simple form of breeding \cite{Holland1975}: a single split point is chosen randomly and the genes are taken from  one parent up to and including the splice point, those of the other after it. The probability that an agent is chosen as a parent is proportional to $20+19(1-t_f)/20$ where $t_f$ is the time step at which the group the agent belongs to became fit. If it never became fit $t_f$ is set to 21. Thus, for example, an agent which belongs to a fit group at the start is 20 times more likely to be selected as a parent than an agent which never belongs to a fit group. In practice the selection works by first selecting a group and then randomly selecting an agent in the group; this is noted to emphasize that the selection process is blind to the number of \1 genes an individual agent has except in so far as it belongs to fit group. The two parents are not required to belong to the same group, in fact, changing that detail does not seem to have any effect. Initially each gene has a probability $p=0.3$ of being \1; this is chosen so that for 256 groups the number of fit groups is one on average. 

\begin{figure}emphasize
\begin{center}
\begin{tabular}{ll}
\textbf{A}&\textbf{B}\\
\includegraphics[width=0.52\textwidth]{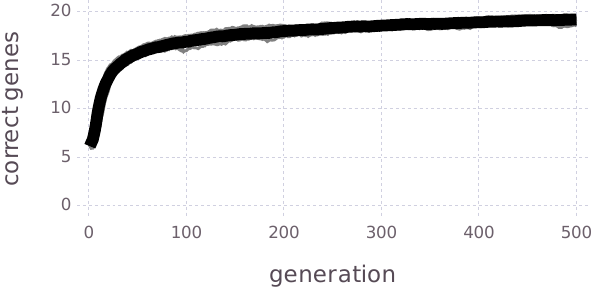}&\includegraphics[width=0.39\textwidth]{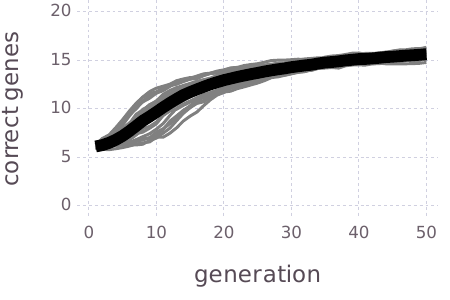}\\
\textbf{C}&\textbf{D}\\
\includegraphics[width=0.52\textwidth]{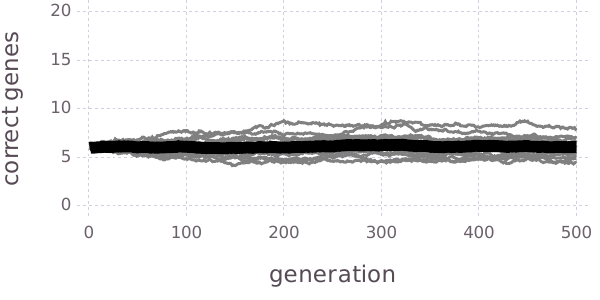}&\includegraphics[width=0.39\textwidth]{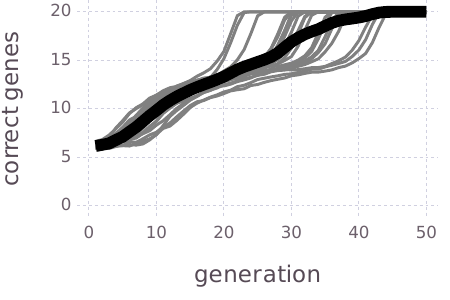}
\end{tabular}
\end{center}
\caption{\textbf{The benefit of cooperation}. Each of these graphs shows the average number of \1 genes for the population of agents as a function of generation number. In each case the simulation has been run 20 times, the thin grey lines correspond to individual trials, the thick black line is the average. In \textbf{A} and \textbf{B} the model is the same, with cooperation across groups of four and fitness of the group giving a preference in breeding; they differ only in the number of generations the simulations run for. In \textbf{C} there is no cooperation and so the fit configuration is lost in the combinatorial forest. In \textbf{D} there is an additional benefit, corresponding to an increased likelihood of being selected for breeding, to an individual who is fit.}
\label{fig:results}
\end{figure}

In Fig.~\ref{fig:results} the result is plotted for 500 (Fig.~\ref{fig:results}\textbf{A}) and 50 (Fig.~\ref{fig:results}\textbf{B}) generations; for comparison the result is also plotted without cooperation in Fig.~\ref{fig:results}\textbf{C}. This demonstrates that the mechanism works, the number of \1 genes rapidly rises to 16 and then more slowly to a value near 20. With 16 \1 genes a group has a probability of 0.97 of being fit, so there is little evolutionary pressure towards completing the set. It is be possible to add some benefit to the agent that is fit without cooperation; the gap between 16 and 20 does not present the formidable combinatorial challenge presented by the gap between six and 20. As a very simple illustration of this in Fig.~\ref{fig:results}\textbf{D} an additional parent selection mechanism has been included, if a group selected to provide a parent contains any fit agents one of these is selected as the parent, otherwise any member of the group is selected with equal probability as before. In this case the number of \1 genes rapidly rises to 20.

\section{Discussion}

Characteristics that require co-adaptation of many characteristics before they become useful are difficult to discover through evolutionary search; in the Baldwin Effect learning bridges the gap between a partial and therefore otherwise useless configuration and the complete and therefore beneficial configuration. In this commentary this is extended from learning to cooperation. Of course, it goes without saying that cooperation can provide an evolutionary benefit and the evolution of cooperation is well studied \cite{AxelrodHamilton1981,Dugatkin1997,Hammerstein2003}: the focus here is slightly different, it is demonstrated that an additonal benefit of cooperation is that it can guide evolutionary towards complex multi-gene characteristics. 

The simulation provided is simple to the point of caricature. However, it is not clear that adding any complexity would make it more realistic because the gulf between any realistic situation and what can sensibly be simulated is so great: the simulation is intended as the simplest possible illustration of the proposed effect, a demonstration rather than any sort of proof. The simulation is very closely based on the simulation provided in \cite{HintonNowlan1987}, the paper that is the focus of this commentary. However, the simulation here omits the learning mechanism that is the core message of that paper. This does not mean that the mechanism proposed here is offered an alternative to that one, there is no reason why they cannot operate in parallel.

The obvious question is `what sort of characteristic might this proposal apply to?'. This is not straightforward, the proposal does rely on there being a benefit to cooperation between individuals with different components of the advantageous characteristic. Certainly, taking the usual cartoon example, having five percent of an eye is useless, \cite{Gould1980}, and this cannot be addressed cooperatively, just as it cannot be addressed by learning. This proposal is more likely to be useful for behavioral and cognitive characteristics, the sort of characteristic of potential benefit to animals capable of cooperation. Indeed, a richer ability to cooperate may be one of the characteristic than evolution can be guided towards though cooperation.

\section*{Code availability} 

Code is available at \texttt{github.com/BaldwinEffect/2024\_cooperation}.

The simulations used \texttt{Julia v1.10.3} along with the \texttt{CSV v0.10.12}, \texttt{DataFrames v1.6.1}, \texttt{Distributions v0.25.107}  and \texttt{GadFly v1.4.0} libraries.

\end{document}